# A multipair-free source of entangled photons in the solid state

Julia Neuwirth[1,*], Francesco Basso Basset[1,*], Michele B. Rota[1], Jan-Gabriel Hartel[2], Marc Sartison[2], Saimon F. Covre da Silva[3], Klaus D. Jöns[2], Armando Rastelli[3], and Rinaldo Trotta[1,*]

[1] Department of Physics, Sapienza University of Rome, Piazzale Aldo Moro 5, 00185 Rome, Italy

[2] Institute for Photonic Quantum Systems, Center for Optoelectronics and Photonics, and Department of Physics, Paderborn University, Warburger Straße 100, 33098 Paderborn, Germany

[3] Institute of Semiconductor and Solid State Physics, Johannes Kepler University Linz, Altenbergerstraße 69, 4040 Linz, Austria

E-mail: Julia.neuwirth@uniroma1.it, Francesco.bassobasset@uniroma1.it, Rinaldo.trotta@uniroma1.it

**Unwanted multiphoton emission commonly reduces the degree of entanglement of photons generated by non-classical light sources and, in turn, hampers their exploitation in quantum information science and technology. Quantum emitters have the potential to overcome this hurdle but, so far, the effect of multiphoton emission on the quality of entanglement has never been addressed in detail. Here, we tackle this challenge using photon pairs from a resonantly-driven quantum dot and comparing quantum state tomography and second-order coherence measurements as a function of the excitation power. We observe that the relative (absolute) multiphoton emission probability is as low as $p_m$ = (5.6 ± 0.6)·10$^{-4}$ ($p_2$ = (1.5 ± 0.3)·10$^{-6}$) at the maximum source brightness, values that lead to a negligible effect on the degree of entanglement. In stark contrast with probabilistic sources of entangled photons, we also demonstrate that the multiphoton emission probability and the degree of entanglement remain practically unchanged against the excitation power for multiple Rabi cycles, despite we clearly observe oscillations in the second-order coherence measurements. Our results, explained by a theoretical model that we develop to estimate the actual multiphoton contribution in the two-photon density matrix, highlight that quantum dots can be regarded as a multipair-free source of entangled photons in the solid state.**

Entangled photon sources find numerous quantum computation and communication applications and are key ingredients for developing photonic quantum networks [1]. The state-of-the-art is currently represented by non-linear crystals, which can generate nearly-maximally entangled photons via the well-known process of spontaneous parametric-down-conversion (SPDC) [2]. However, SPDC sources are not ideal for practical applications. The intrinsic property of being a probabilistic photon-source gives SPDC a non-zero probability of emitting more than one entangled photon pair per excitation pulse [3]. Thus, when enhancing the brightness of the sources, the multipair emission probability also increases, impairing the entanglement fidelity [4]. In turn, this limits the technological potential of SPDCs, for example, it reduces the maximum achievable secure key rate in quantum key distribution [5,6] and hampers the scalability of multiple photon experiments [2]. Solid-state-based quantum emitters, notably epitaxial quantum dots (QDs), have the potential to overcome this hurdle and promise near-deterministic generation of strongly entangled photons [7,8] with, in principle, no compromise between multiphoton emission and brightness [9,10]. However, recent studies on coherently driven QDs have highlighted that re-excitation processes can lead to non-negligible values of the second-order autocorrelation function $g^{(2)}(0)$ (as measured via a Hanbury Brown and Twiss (HBT) interferometer) [11,12], which would negatively impact the level of entanglement of the emitted photons. On the one hand, there is experimental evidence that the degree of entanglement depends on the finite values of $g^{(2)}(0)$ [13]. On the other hand, these experiments do not employ resonant excitation schemes [11], and it is often experimentally challenging to ascertain whether the entanglement degradation and the finite $g^{(2)}(0)$ values are due to true multiphoton emission or background light originating from the excitation laser and/or other states not involved in the entangled photon generation process, especially at high excitation powers. The matter is further complicated by inconsistencies in the literature [13–15] on how to relate the information on multipair emission given from the $g^{(2)}(0)$ to the polarization density matrix. Therefore, it remains unclear whether QDs can be regarded as a multipair-free source of entangled photons. This letter provides a positive answer to this question by carefully looking at the interplay between second-order coherence, multiphoton emission, and the degree of entanglement in resonantly driven QDs.

We use GaAs/AlGaAs QDs grown by droplet etching epitaxy [16,17]. This class of QDs provides state-of-the-art values of fidelity to a maximally entangled state without resorting to spectral or temporal selection [14]. The sample design, as described in Ref. [18], uses a distributed Bragg reflector planar cavity and a solid immersion lens to achieve an extraction efficiency at the first lens of about 10%. The sample is operated at 4 K in a closed-cycle cryostat. It is driven by a Ti:sapphire laser with a repetition rate of 80 MHz and a pulse duration—adapted with a pulse-slicer—of approximately 10 ps. The laser is tuned to half the energy difference between the biexciton (XX) and the ground state (0) to achieve resonant two-photon excitation [19,20], as illustrated in the energy scheme in Fig. 1(a). The mismatch between the laser energy and the emission energies of the exciton (X) and the XX state, due to a XX binding energy of 3.97 meV (1.79 nm in wavelength), allows for efficient spectral filtering of the laser back-reflected light. Emission spectra for two different pump powers and pair generation rates are shown in Fig. 1(b). The two peaks with higher intensity correspond to the two transitions of the XX-X cascade, whereas the secondary peak (X*) is unrelated to the cascade and has a linear dependence on the laser pulse area [21]. (X* is most probably a negatively charged exciton weakly excited through its high-energy acoustic phonon sideband). Focusing on the X and XX lines, Fig. 1(c) shows a typical excitation power dependence of their integrated intensity, with Rabi-oscillations reflecting the typical coherent excitation behavior of a three-level system [20]. It is worth mentioning that the emission intensity was optimized using an additional off-resonant light field to accelerate charge fluctuations and maximize the "on"-time of the QD [14]. Finally, we selected a QD with a fine structure splitting lower than the spectral resolution of our measurement system of 0.5 µeV. This minimizes the induced relative phase precession in the polarization-entangled state [22].

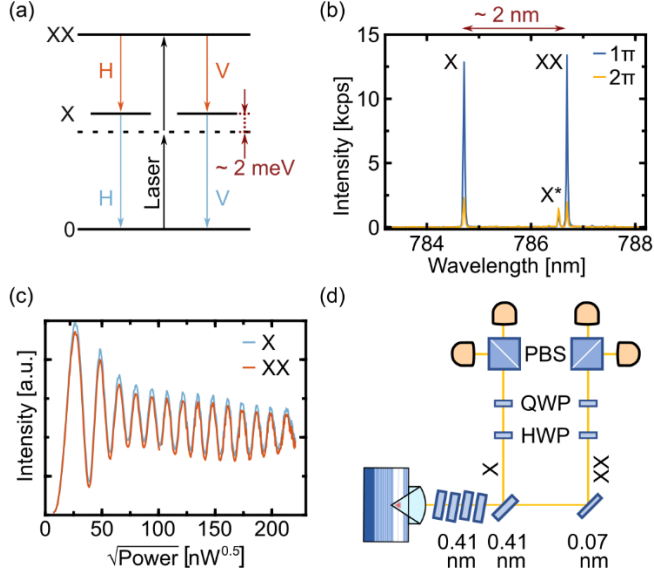

Figure 1 (a) Energy scheme of the XX-X cascade, including the two-photon excitation scheme. The QD is excited to the XX state by a pulsed laser tuned to half the energy difference between the XX level and the ground state 0. The two-photon cascade via the X state emits two polarization-entangled photons. (b) Two-photon photoluminescence spectrum of the QD at π and 2π-pulse. An additional line (X*) unrelated to the cascade is observed. (c) Rabi oscillation of the X and XX emission intensity vs laser power. (d) Quantum state tomography and second-order auto-correlation measurement setup. The laser is filtered by a set of volume Bragg gratings. The X and XX are then spectrally filtered with bandwidths of 0.41 and 0.07 nm, respectively. For the auto-correlation measurements, the PBSs are exchanged with 50:50 BSs.

A simple sketch of the measurement setup is illustrated in Fig 1(d). A set of volume Bragg gratings with a bandwidth of 0.41 nm are placed to remove scattered laser light and background emission. The X and XX are spectrally separated using filters with bandwidths of 0.41 and 0.07 nm, respectively. Note that the XX filter bandwidth is chosen narrower to remove the undesired side peak X*, but both bandwidths are large enough not to filter out any significant fraction of the light emitted by the X and XX states. The rest of the setup performs polarization-resolved cross-correlation measurements between XX and X photons. This is built up, for both the X and XX detection paths, by a combination of a half-waveplate and a quarter-waveplate (for state rotation), a polarizing beam splitter (PBS, for state projection), and two avalanche photodiodes (APDs). When measuring the second-order correlation function $g^{(2)}(\tau)$ of the X and XX emission, the same setup is repurposed by replacing the PBSs with two 50/50 beam splitters (BSs). This configuration results in a standard HBT interferometer.

Using the HBT setup, we measured $g^{(2)}(\tau)$ for different excitation powers to investigate how the photon generation rate is linked to the multi-pair emission. Fig. 2(a) shows, in logarithmic scale, the coincidence histograms for the X line at the first maximum and minimum in the Rabi oscillations, namely at π- and 2π-pulse area. Both pump regimes result in heavily suppressed coincidences around zero-time delay, indicating a nearly single-pair emission behavior. From the coincidence histograms we can now calculate the $g_X^{(2)}(0)$. The $g_X^{(2)}(0)$ is defined as the coincidences of photons generated with the same laser pulse (zero-time delay), normalized by that of an equally bright Poisson distributed source.

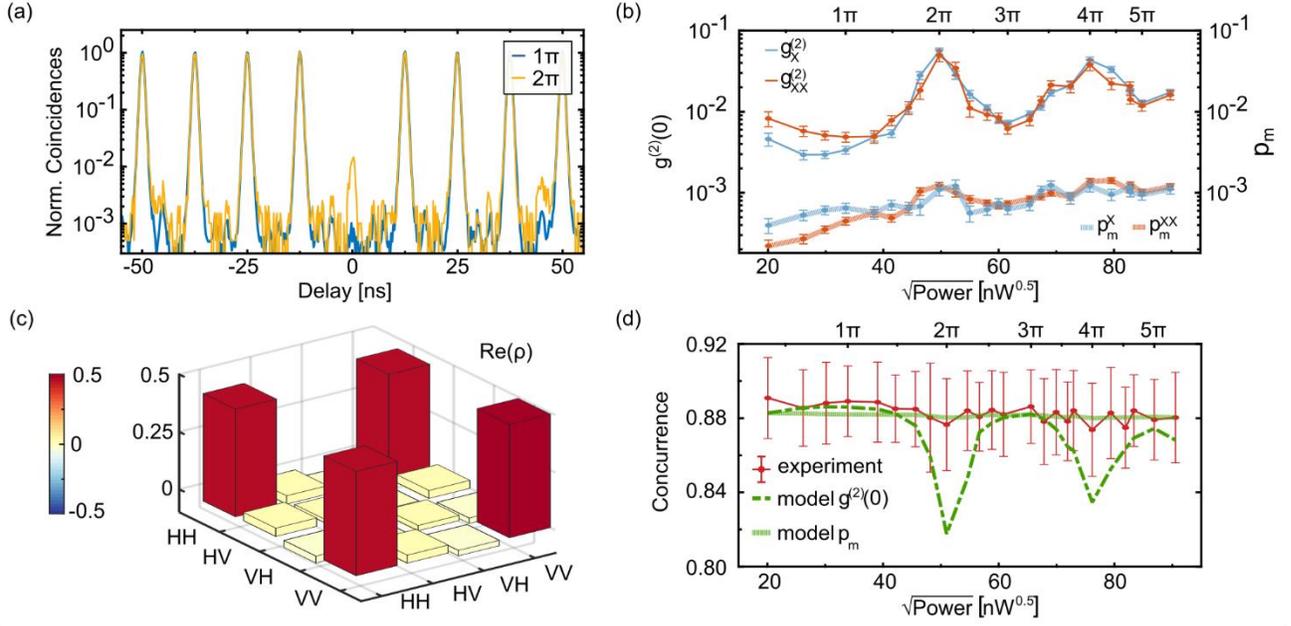

*Figure 2(a) Second-order auto-correlation function $\tilde{g}_X^{(2)}(\tau)$, normalized to the side peaks (12.5 ns), measured at π and 2π-pulse area. (b) Measured $g^{(2)}(0)$ and multiphoton probability $p_m$ from the X and XX lines for different excitation powers. The $g^{(2)}(0)$ for both the X and XX show an increase at even π-pulse areas in contrast to $p_m$. (c) Real part of the density matrix of the XX-X photon polarization state measured at π-pulse area. (d) Concurrence for different excitation powers. The experimental data (red) are compared with both the predictions of Eq. 1-2 (dashed green line) and our model using Eq. 5 (continuous green line). The error bars in the modeled concurrences, which enter by the measured multiphoton probability and are estimated by a Monte Carlo simulation, are within the line thickness.*

For the pump power at π-pulse area, which corresponds to the highest pair generation rate and thus to the most usual QD driving condition, the measured $g_X^{(2)}(0)$ is (3.4 ± 0.4)·10⁻³. To directly compare this value with the literature, we need to normalize the coincidences at zero-time delay to the coincidences at consecutive excitation laser pulses (12.5 ns time delay), as done in Fig. 2(a). This leads to $\tilde{g}_X^{(2)}(0) = \eta_{blink} g_X^{(2)}(0)$ of = (1.0 ± 0.1)·10⁻³, which is lower by a factor $\eta_{blink} = 0.29$, attributed to the blinking of the source, an intermittent emission behavior of the QD (see the Supplementary Materials for further information). We emphasize that $\tilde{g}_X^{(2)}(0)$ is not the correct quantity to estimate the second-order correlation function at zero-time delay. Still, it is commonly used in the literature. The value we obtain is very similar to those reported for the GaAs/AlGaAs system [23], only surpassed by the record value obtained using single-photon detectors with ultralow dark count rates [21]. However, we emphasize that the figures reported here are achieved without polarization suppression of the back-reflected excitation laser or post-selection schemes. The $g_X^{(2)}(0)$ at 2π-pulse area excitation is (55.9 ± 5.1)·10⁻³ ($\tilde{g}_X^{(2)}(0)$ =(17.1 ± 1.5)·10⁻³), about 15 times higher than the value obtained under π-pulse conditions, but still well within the anti-bunching regime. To better understand this behavior, we analyze the complete power dependence for both the X and XX emissions. Fig. 2(b) shows how $g_X^{(2)}(0)$ behaves with power up to above 5π-pulse area, as from the previously measured Rabi oscillations (note that the measurements in Fig. 1(c) are performed on a different QD). The values obtained for the X and XX lines are the same within the margin of error, which is estimated by Gaussian error propagation and assuming a Poisson distribution of the measured coincidence counts. Clear oscillations are observed at even numbers of π-pulse areas, as shown in Fig. 2(b). Similar oscillations have been experimentally observed in QDs [11], though only in resonantly driven two-level systems (2LS), in which the oscillation stems from an increased multiphoton emission at even π-pulse areas. Specifically, the 2LS can spontaneously decay to the ground state instead of undergoing an even number of π rotations. In this case, if the excitation laser pulse is still present, a second excitation is possible [11]. This breaks the even π-pulse area excitation into two uneven π-pulse area excitations with a radiative recombination in-between. While for cascaded quantum ladder systems, re-excitation is expected to be strongly suppressed [21] - as it can take place only when both the X and XX photons are emitted - the oscillations observed in Fig. 2(b) may indicate that the multiphoton emission probability is non-negligible and oscillates with power. To show that this is not the case, we use an approach that does not rely on any assumption on the physical origin of the multiphoton emission. Instead, we infer its contribution from the experimental $g^{(2)}(0)$ of the photons emitted by the XX-X cascade and investigate the effect of its finite values on the measured degree of entanglement, as discussed below.

We perform polarization-resolved XX-X cross-correlation measurements to reconstruct the two-photon density matrix using the setup from Fig. 1(d). Rotations in the polarization state induced by the optical components in the setup are compensated using a set of linear waveplates to maximize the fidelity to the expected Bell state $|\phi^+\rangle$ [24]. The density matrix is reconstructed from a set of 36 correlation measurements using quantum state tomography and maximum likelihood estimations [25]. Fig. 2(c) shows the resulting real part of the density matrix at π-pulse area. The imaginary part does not contain significant terms (no matrix element is above 0.045 in absolute value). Furthermore, the fidelity to $|\phi^+\rangle$ is (0.93 ± 0.01), while the concurrence is (0.89 ± 0.02), values that are comparable with the literature [23] and bested only by those obtained with strain-tunable QDs [14]. We note that the influence of non-measurable fine structure splitting (below 0.5 µeV) should affect the fidelity by less than 1% [24]. Additionally, we characterized the retardance of the waveplates and detector dark counts and simulated their impact on the density matrix estimation to conclude that their impact amounts to less than 0.7% on the Bell state fidelity. The error bars are estimated with a Monte Carlo method assuming a Poisson distribution of the measured coincidence counts. Quite remarkably, no significant variation of these figures of merits is observed at 2π-pulse area and across the whole range of powers investigated in this work, as shown in Fig. 2(d). On the one hand, this result is in stark contrast with the behavior observed in entangled photon sources relying on SPDC [2], thus reflecting the intrinsically different nature of the photon generation processes in QDs compared to non-linear crystals. On the other hand, this result also appears in contrast with the common idea that multiphoton emission, whose presence could be associated with the non-zero values in the $g_X^{(2)}(0)$ and $g_{XX}^{(2)}(0)$ measurements, reported in Fig. 2(b), affects the degree of entanglement. Specifically, one expects multipair emission to degrade the measured entanglement because of erroneous detection events of an X and XX photon belonging to two different photon pairs stemming from subsequent (and thus uncorrelated) cascades. The effect of multiphoton components has been included in the density matrix $\rho$ in previous works using knowledge of the $g_{X/XX}^{(2)}(0)$ in the following way [13,14]:

$$\rho = \frac{1}{4}(1-k)\mathbb{1} + k\rho_0 \qquad (1)$$

$$1-k = \frac{1}{2}\left(g_X^{(2)}(0) + g_{XX}^{(2)}(0)\right) \qquad (2)$$

with $\rho_0$ being the density matrix neglecting accidentals coincidences due to multiphoton components, $k$ being the fraction of photon pairs that come from a radiative XX-X cascade with respect to the total number of detected pairs. According to this model, a link between the concurrence and the XX (or X) population should be visible, as indicated by the dashed line in Fig. 2(d), obtained using equations (1)-(2) in combination with the $g_X^{(2)}(0)$ and $g_{XX}^{(2)}(0)$ measurements reported in Fig. 2(b). We notice that the discrepancy is evident at even π-pulses, which motivates why it is important to investigate the entanglement degree in a wide range of excitation powers. A clear inconsistency between the measurements and this commonly used theory is observed. This raises two questions: can the $g^{(2)}(0)$, as measured via HBT, actually help to estimate the effect of multiphoton emission on the density matrix? Is multi-photon emission present in our system? We address the matter using a probability-based model to estimate how multiphoton emission affects the measurements of $g^{(2)}(0)$ and $\rho$.

We define the parameter $p_m$ as the probability that a successful cascaded photon emission is followed by re-excitation and a second cascaded photon emission. For simplicity, we exclude multiple emission events beyond double, which are expected to be a negligible fraction in all realistic cases. $p_m$ is related to the photon generation distribution, as $p_m = p_2/(p_1 + p_2)$ where $p_1$ ($p_2$) is the probability per excitation pulse of generating a single (two) photon per pair per transition. It is worth pointing out that while the $g^{(2)}(0)$ is directly related to $p_m$, it also depends on the probability of photon generation at the source. This fact is well known from estimations of upper bounds for multiphoton emission from intensity autocorrelation measurements [26,27]. In our case, given the overall efficiency of the setup well below unity, we estimate that

$$g^{(2)}(0) = \frac{2p_m}{\eta_{blink}\eta_{prep}(1+p_m)^2} \quad (3)$$

The complete derivation of equation (3) is discussed in the Supplemental Material. Here we have split the probability of photon generation in the blinking factor $\eta_{blink}$ and the preparation fidelity $\eta_{prep}$, which is defined as the probability that a laser pulse will result in the emission of a photon pair if the QD is in its active ground state.

We use the same probability-based method to estimate $k$, the fraction of photon pairs that originate exclusively from the dot, needed to estimate the density matrix [22]. In the limit of a small overall efficiency of the setup (the complete formula is reported in the Supplemental Material), we have

$$1 - k = \frac{2p_m}{1+3p_m} \quad (4)$$

Here the numerator is proportional to the probability of having a coincidence between a photon emitted by a first XX-X cascade and a second photon coming from a re-excitation event, divided by all the possible coincidence combinations, including the ones of entangled photons coming from the same cascade.

We immediately notice that while the $g^{(2)}(0)$ is inversely proportional to $\eta_{prep}$, this is not the case for the quantity $k$. An intuitive explanation can account for this difference: In autocorrelation measurements, the zero-time delay coincidences are proportional to $\eta_{prep}$ and $p_m$, whereas the coincidence peaks used for normalization are instead proportional to $(\eta_{prep})^2$. This is related to the fact that the two different laser pulses must successfully excite the QD. In the case of the density matrix measurements, one is instead only interested in cross-correlation coincidences at zero-time delay, without normalization on peaks at other time delay. This is the reason why the additional term due to $\eta_{prep}$ is not present in Eq. (4), which instead contains only $p_m$.

Combining the two formulas makes it possible to estimate the effect of multiphoton emission on the two-photon density matrix based on the knowledge of $g^{(2)}(0)$. If we consider for simplicity the limit of a small fraction of multiphoton emission events, which is an excellent assumption for our source, we obtain the following expression:

$$1 - k \approx \frac{g_X^{(2)}(0) + g_{XX}^{(2)}(0)}{2}\eta_{prep}\eta_{blink} = \frac{\tilde{g}_X^{(2)}(0) + \tilde{g}_{XX}^{(2)}(0)}{2}\eta_{prep} \qquad (5)$$

Note that our model is an extension of the formula reported in Ref. [15], which was an approximation derived for generic noise in the auto-correlation measurement. We underlined that Eq. (5) can also account for different normalization criteria of the second-order correlation function by introducing the appropriate corrective factor.

To verify the model, we also need to estimate $\eta_{prep}$. Since $\eta_{prep}$ strongly varies across Rabi oscillations, studying the whole power dependence and in particular the even π-pulse areas, allows us to investigate the role of this variable. This quantity is often inferred from the power dependence of the photoluminescence as in Fig. 1(c), using a model to extract the occupation number of the XX state. Here we opt for a different approach, which requires fewer assumptions and prevents from neglecting contributions due to power-dependent blinking dynamics. We estimate the preparation fidelity from the normalized zero-time delay coincidences in an intensity cross-correlation measurement between the X and XX photons [15,28]. This is conducted with the same setup as illustrated in Fig. 1(d) without any polarization selection. The XX-X cross-correlations histograms, normalized to the side peaks (12.5 ns), for π- and 2π-pulse areas are shown in Fig. 3(a). The peaks at zero-time delay, due to photons belonging to the same cascade, are compared to the neighboring side peaks from coincidences of photons belonging to subsequent excitation pulses. $\eta_{prep}$ as a function of the pump power is reported in Fig. 3(b), displaying the expected population oscillations, with the highest (lowest) value of 0.93 (0.14) at π- (2π-) pulse area. The highest preparation fidelity is comparable to those published in the literature for similar GaAs/AlGaAs QDs [6]. As anticipated, we highlight that this method can access the occupation number of the biexciton state under coherent excitation more directly than the most used approach of monitoring how the photoluminescence intensity varies with excitation power, as explained in [13] and the corresponding Supplementary Material. First, it does not require any assumption on how to model the damping of the Rabi oscillations. Second, power-dependent changes of the "on"-time fraction of the QD due to blinking also affect the photoluminescence intensity. They should not be neglected when inferring information about the coherent coupling between the ground level and the XX state. In our investigated QD, the blinking dynamics show a significant power dependence of its characteristic time [29], ranging from 16.1 μs to 1.6 μs (see Supplementary Material). However, while applying an additional off-resonant light field achieving partial photo-neutralization, $\eta_{blink}$ is approximately 0.3 in almost the entire investigated range of pump powers.

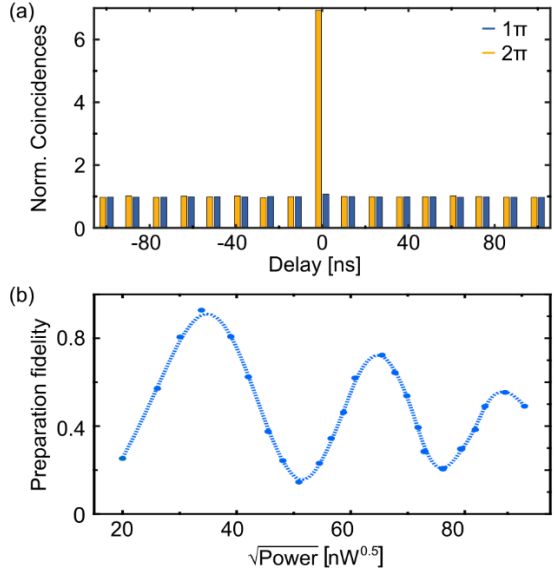

*Figure 3 (a) Coincidence histograms between X and XX photons, normalized to the side peaks (12.5 ns), for π and 2π-pulse area excitation. (b) Measured preparation fidelity (dots) for different pumping regimes, with a spline interpolation to visualize the oscillating behavior as a guide to the eye. The error bars are within the dot size.*

Given these findings, it is apparent from Eq. (3) that the oscillations in the $g^{(2)}(0)$ are not necessarily related to variations in the multiphoton emission probability but rather to oscillations in the preparation fidelity. Using the measured values of $g^{(2)}(0)$, $\eta_{prep}$, and $\eta_{blink}$, Eq. (3) can be exploited to calculate $p_m$ directly. This approach does not rely on any specific assumption on the physical origin of the multiphoton component, yet it allows us to quantify its impact on the density matrix. The result of this procedure, which is shown as the continuous line in Fig. 2(b), clearly highlights that the fraction of emission events related to multiphoton emission does not vary noticeably with the pump power, i.e., $p_m$ barely varies with the pulse area. This, in combination with Eq. (4), readily explains why the oscillations in the $g^{(2)}(0)$ are not linked to any oscillation of the degree of entanglement with pump power. Moreover, its quantitative contribution appears negligible, as we can estimate that the level of multiphoton emission measured at maximum source brightness (π-pulse area) is $p_m$= (5.6 ± 0.6)·$10^{-4}$ corresponding to an absolute value of $p_2$= (1.5 ± 0.3)·$10^{-6}$. These values applied to an ideal Bell state would result in a concurrence of 99.8%. The lower concurrence measured in our experiment is attributed to cross dephasing mechanisms in the bright exciton state for this particular QD [15]. Ultimately, using Eq. (5) in combination with Eq. (1)-(2), where $\rho_0$ is the density matrix measured at minimum excitation power, we obtain an excellent agreement between our model and the experimental data for the concurrence reported in Fig. 2(d).

In conclusion, we have shown that the effect of multiphoton emission on the degree of entanglement of photons emitted by resonantly-driven QDs is negligible and, contrarily with the behavior reported for single-photon generation in two-level systems [30], does not vary significantly with pump power. This occurs despite we observe oscillations in the values of the $g^{(2)}(0)$ by more than one order of magnitude. We illustrate that the variations of the $g^{(2)}(0)$ with pump power are not necessarily related to the variation of the multiphoton emission probability, but rather to variations of the preparation fidelity of the excited state. With the support of a probability-based model, we identify the actual contribution of the relative multiphoton emission $p_m$, as estimated from the $g^{(2)}(0)$, which enters the simulation of the quantum tomography results and successfully reproduces the experimental data.

This work thus tackles a fundamental obstacle for state-of-the-art entangled photon sources based on SPDC: the relationship between pump power, brightness, and entanglement quality. Even though the absence of a tradeoff between brightness and entanglement due to the multiphoton emission has long been a motivation for developing QD-based sources, we finally provide a thorough experimental study demonstrating that multiphoton emission is negligible and does not negatively affect the generated entangled states across a wide range of excitation powers. The result strengthens the case for QDs providing highly entangled photons for complex quantum information protocols. The entanglement fidelity reported in this work was yet below unity, an evidence that is attributed to residual decoherence mechanisms between the bright exciton states in the QD system [14]. This effect, which can be significantly lower in selected QDs [14,31], could be further reduced with the help of photonic cavities to shorten the lifetime of the optical transition [32,33]. This strategy is predicted to increase the entanglement fidelity above 0.99 and bring QD entanglement to the same level as SPDC sources [34–36]. Since this performance is achieved without sacrificing brightness due to multiphoton emission, QD-based entangled photon sources will be an essential element for performing quantum information protocols of ever-increasing complexity.

**Acknowledgements**

This project has received funding from the European Research Council (ERC) under the European Union's Horizon 2020 Research and Innovation Program under Grant Agreement No. 679183 (SPQRel), by the European Union's Horizon 2020 Research and Innovation Program under Grant Agreement No. 899814 (Qurope) and No. 871130 (Ascent+), by the Austrian Science Fund via the Research Group FG5 and I 3762, by MIUR (Ministero dell'Istruzione, dell'Universita e della Ricerca) via project PRIN 2017 Taming complexity via Quantum Strategies a Hybrid Integrated Photonic approach (QUSHIP) No 2017SRNBRK.